\documentclass[universe,article,accept,moreauthors]{Definitions/mdpi}
\usepackage{amsmath}
\usepackage{amsfonts}
\usepackage{amssymb,bm}
\usepackage{siunitx}
\usepackage{color}

\firstpage{1}
\pubvolume{xx}
\issuenum{1}
\articlenumber{5}
\pubyear{2019}
\copyrightyear{2019}
\history{Received: date; Accepted: date; Published: date}

\Title{The State of the Art in Constraining Axion-to-Nucleon Coupling and Non-Newtonian
Gravity from Laboratory Experiments}

\Author{Vladimir M. Mostepanenko $^{1,2,3}$ and Galina L. Klimchitskaya $^{1,2}$
}

\address{%
$^{1}$ \quad Central Astronomical Observatory at Pulkovo of the Russian Academy of Sciences, Saint Petersburg, 196140, Russia\\
$^{2}$ \quad Institute of Physics, Nanotechnology and
Telecommunications, Peter the Great Saint Petersburg
Polytechnic University, Saint Petersburg, 195251, Russia\\
$^{3}$ \quad Kazan Federal University, Kazan, 420008, Russia}

\corres{Correspondence: vmostepa@gmail.com}

\abstract{Constraints on the Yukawa-type corrections to Newton's
gravitational law and on the coupling constant of axionlike particles
to nucleons obtained from different laboratory experiments are reviewed
and compared. The constraints on non-Newtonian gravity under discussion
cover the wide interaction range from nanometers to millimeters and
follow from the experiments on neutron scattering, measuring the
Casimir force and Cavendish-type experiments. The constraints on the
axion-to-nucleon coupling constant following from the magnetometer
measurements, Cavendish-type experiments, Casimir physics, and
experiments with beams of molecular hydrogen are considered which
refer to the region of axion masses from $10^{-10}$ eV to 200 eV.
Particular attention is given to the recent constraints obtained
from measuring the Casimir force at nanometer separation distance
between the test bodies. Several proposed experiments focussed on
constraining the non-Newtonian gravity, axionlike particles and
other hypothetical weakly interacting particles, such as chameleons
and symmetrons, are discussed.
}


\begin{document}
\section{Introduction}

In recent years, more and more experiments are devoted to a search
for some uncharged weakly interacting elementary particles which are
not accessible for the accelerator techniques used in high energy
physics. Many particles of this kind were predicted in different
extensions of the Standard Model, supergravity and supersymmetry
\cite{1,2}, and their discovery {  would} have a profound effect on
several branches of physics and change our picture of the world.
Thus, an exchange of light scalars (for instance, dilatons \cite {3})
between atoms of neighboring macrobodies results in the Yukawa-type
correction to the Newtonian gravitational law and opens up a new
area to the experimenter: the search for non-Newtonian gravity
\cite{4}. The predicted light pseudoscalar particles called axions
\cite{5,6} not only provide a plausible explanation for the problem
of strong $CP$ violation and large electric dipole moment of a neutron
in QCD, but suggest themselves as possible constituents of dark
matter \cite{7}.

The Yukawa-type interaction at short separations between the test
masses has long been searched for as the so-called {\it fifth force}
in the gravitational experiments of E\"{o}tvos and Cavendish-type
\cite{4}. According to the results of \cite{8,9}, both the Yukawa-
and power-type corrections to Newtonian gravity can be recorded also
in experiments on measuring the van der Waals and Casimir forces
caused by the zero-point and thermal fluctuations of the electromagnetic
field. The development of extradimensional theories, which anticipate
the compactification scale at energies of the order of 1~TeV
\cite{10,11}, made one more prediction of the Yukawa-type corrections
to Newton's law at short separations \cite{12,13} and stimulated their
search.

An exchange of one pseudoscalar particle between two fermions results
in the spin-dependent interaction potential \cite{14,15} which turns
into zero after an averaging over the volumes of two unpolarized
macroscopic bodies. Because of this, the process of one-axion exchange
does not lead to any fifth force. An effective potential due to a
simultaneous exchange of two axions is more complicated and depends
on the details of an axion-to-fermion interaction. If this interaction
is described by the pseudoscalar Lagrangian, the spin-independent
effective potential leading to some fifth force between two macroscopic
bodies is obtained \cite{14,15}. If, however, an axion-to-fermion
interaction is described by the pseudovector Lagrangian, as it holds
for QCD axions which are the pseudo-Nambu-Goldstone bosons, the
character of an effective potential due to the exchange of two axions
remains a mystery \cite{16}.

The constraints on the constant of Yukawa-type interaction $\alpha$ are
usually obtained as functions of the interaction range $\lambda=M^{-1}$,
where $M$ is a mass of the exchange scalar particle (here and below we
put $\hbar=c=1$). For non-Newtonian gravity originating from extra
dimensions $\lambda$ has the meaning of the compactification scale.
Depending on the interaction range, the strongest constraints on $\alpha$
follow from different experiments. Over a long period of time, experiments
on measuring the Casimir force were used for obtaining the strongest
constraints on $\alpha$ over the wide interaction range below a few
micrometers \cite{17}. For larger $\lambda$ up to a few centimeters the
strongest constraints on $\alpha$ were found from checking the inverse
square law of Newtonian gravity (the Cavendish-type experiments
\cite{18,19,20}) and for even greater $\lambda$ from tests of the
equivalence principle (the E\"{o}tvos-type experiments \cite{21,22}).
At the moment, the strongest constraints on $\alpha$ in the interaction
range below approximately 10~nm follow from the experiments on
neutron scattering \cite{23,24,25}. The Casimir force measurements
(and their modification where the Casimir force is nullified \cite{26})
allow obtaining the strongest constraints on $\alpha$ within the
interaction range from approximately $\lambda=10~$nm to $\lambda=8~\mu$m.
For larger $\lambda$ the gravitational experiments preserve their
leading role in constraining the Yukawa-type corrections to Newtonian
gravity.

Constraining the parameters of axions and axionlike particles is the
subject of extensive studies. There are many types of axions in the
literature grouped in the models of hadronic (QCD) axions \cite{27,28}
and Grand Unified Theory (GUT) axions \cite{29,30} which in turn are
divided into many submodels. Axionlike particles can interact with
photons, electrons, and nucleons, and each interaction channel is
used in the laboratory experiments and astrophysical observations
dedicated to the axion quest (see reviews \cite{1,7,31,32,33}). Our
interest here is focussed on the interaction of axions with nucleons.
The constraints on the strength of an axion-to-nucleon interaction
constant $g_{an}$ are usually obtained as functions of the nucleon
mass $m_a$. In fact the strongest constraints on $g_{an}$ follow
from astrophysical arguments on stellar cooling accounted for by
the axion emission \cite{34}. It has been known, however, that the
theory of dense nuclear matter suffers from significant
uncertainties which make the obtained constraints not fully
reliable \cite{34}. At the moment, the strongest laboratory
constraints on $g_{an}$ in the range of of $m_a < 1~\mu$eV follow
from the magnetometer measurements \cite{35} and for heavier axions
from the Cavendish-type \cite{19,36} and Casimir-less \cite{26,37}
experiments. For $m_a > 0.5~$eV the strongest laboratory limits on
$g_{an}$ were obtained from the experiment with beams of molecular
hydrogen \cite{38,39} and for $m_a > 200~$eV from the experiment on
nuclear magnetic resonance \cite{39}
{  (see also review \cite{39a} of the laboratory constraints on
both the axionlike particles and Yukawa-type interactions with a focus
on neutron experiments)}.

In this article, we review constraints on the Yukawa-type interaction
and interaction of axions with nucleons obtained during the last few
years from measurements of the Casimir force. These constraints are
compared with the strongest limits following from the gravitational
and some other laboratory experiments in the same interaction ranges.
We also discuss several proposals focussed on further strengthening
the constraints on both non-Newtonian gravity and axionlike particles.

The article structure is the following In Section 2, we consider
constraints on the strength of Yukawa-type correction to Newtonian
gravity obtained from different measurements of the Casimir force
and Casimir-less experiments. In Section 3, the same is done with
respect to the constant of axion-to-nucleon interaction. Section 4
presents the constraints on both non-Newtonian gravity and axionlike
particles obtained recently from measuring the Casimir force at
nanometer separation range. These constraints are compared with those
outlined in Sections 2 and 3 and with the most strong results
following from some other laboratory experiments. In Section 5,
several proposed experiments are discussed allowing further
strengthening of the obtained constraints. Section 6 contains our
discussion. Finally, in Section 7 the reader will find the
conclusions.

\section{The Yukawa-Type Correction to Newtonian Gravity and Constraints
on It from the Casimir Effect}

The Yukawa-type correction to the Newtonian gravitational potential between
two particles with masses $m_1$ and $m_2$ situated at the points
$\mbox{\boldmath$r$}_1$ and $\mbox{\boldmath$r$}_2$ is conventionally written
in the form \cite{4}
\begin{equation}
V_{\,\rm Yu}(r_{12})=-\alpha\frac{Gm_1m_2}{r_{12}}\,e^{-r_{12}/\lambda},
\label{eq1}
\end{equation}
\noindent
where $r_{12}=|\mbox{\boldmath$r$}_1-\mbox{\boldmath$r$}_2|$,
$\alpha$ is the dimensionless interaction constant of the Yukawa interaction,
and $G$ is the Newtonian gravitational constant.
As mentioned in Section~1, the interaction range $\lambda$ has a meaning of either
the Compton wavelength of an exchange scalar particle of mass $M$,
$\lambda=M^{-1}$, or the compactification scale of extra spatial dimensions.

For two test bodies $V_1$ and $V_2$ at the closest separation $a$ the Yukawa
interaction energy is obtained by interacting (\ref{eq1}) over their volumes
\begin{equation}
E_{\,\rm Yu}(a)=-\alpha G\rho_1\rho_2\int_{V_1}\int_{V_2}d\mbox{\boldmath$r$}_1
d\mbox{\boldmath$r$}_2\frac{e^{-r_{12}/\lambda}}{r_{12}},
\label{eq2}
\end{equation}
\noindent
where  $\rho_1$ and $\rho_2$ are the respective mass densities.
The Yukawa-type force acting between the two bodies can be calculated as
the negative derivative of (\ref{eq2})
\begin{equation}
F_{\,\rm Yu}(a)=-\frac{\partial E_{\,\rm Yu}(a)}{\partial a}.
\label{eq3}
\end{equation}

It should be noted that although the Yukawa-type force is often called
a "correction" to Newtonian gravity, at short separations below a micrometer
such an experimentally allowed "correction" may exceed the gravitational force
by many orders of magnitude. For this reason, when obtaining the constraints,
it is customary to omit the contribution of the genuinely Newtonian gravitation.
This does not make any impact on the obtained results.

We begin with constraints found at the shortest interaction range from measuring
the lateral Casimir force acting between the sinusoidally corrugated surfaces.
It is common knowledge that the Casimir force between two parallel plates or
between a plate and a sphere acts in perpendicular direction to the surfaces.
However, for surfaces covered with the sinusoidal corrugations of a common
period but having some nonzero phase shift $\Phi$ (see Figure~1) the Casimir
free energy ${\cal F}_C(a,\Phi,T)$ depends not only on $a$ and on temperature $T$,
but also on $\Phi$. This leads to initiation of the lateral Casimir force
\begin{equation}
F_{C,\,\rm lat}(a,\Phi,T)=
-\frac{\partial {\cal F}_C(a,\Phi,T)}{\partial \Phi}.
\label{eq4}
\end{equation}
\begin{figure}[!h]
\centering
\vspace*{-8.5cm}
\hspace*{-2cm}
\includegraphics[width=20 cm]{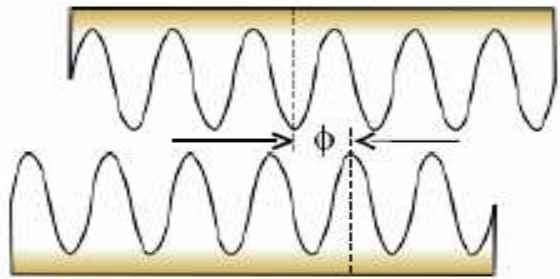}
\vspace*{-16.cm}
\caption{The schematic of two sinusoidally corrugated plates having the phase
shift $\Phi$ between corrugations.\label{fig1}}
\end{figure}

{  The sinusoidally corrugated surfaces are subjected to both the
normal Casimir force acting perpendicular to the surface and the lateral
Casimir force acting in a tangential direction. Due to the presence of
corrugations, the closest separation between the two surfaces is smaller in
this case than for the smooth bodies. Because of this, by measuring both the
lateral and normal forces between corrugated surfaces it becomes possible
to constrain hypothetical interactions within the shortest interaction range.
Here, we discuss the constraints obtained from experiments on measuring both
the lateral and normal Casimir forces starting from the lateral one.
The constraints found from measurements of the normal Casimir force acting
between the smooth surfaces are also considered.}

The lateral Casimir force (\ref{eq4}) was first predicted theoretically in \cite{40}
and observed experimentally in \cite{41,42}. In \cite{43,44} it was measured more
precisely in the configuration of a corrugated sphere in front of a corrugated plate
with increased amplitudes at room temperature.
\begin{figure}[!b]
\centering
\vspace*{-7.5cm}
\hspace*{-2cm}
\includegraphics[width=22 cm]{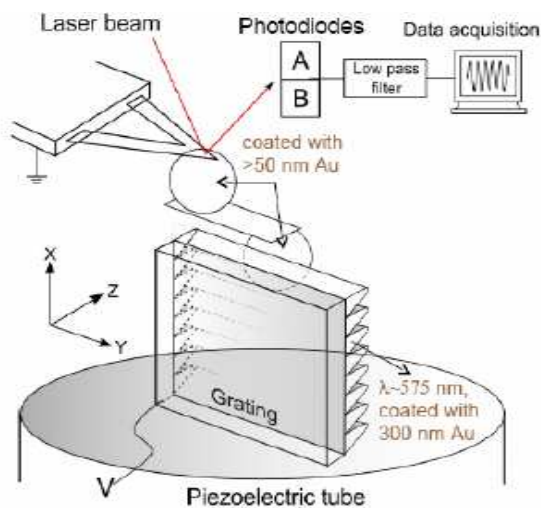}
\vspace*{-16cm}
\caption{Schematic diagram of the experimental setup using an atomic force
microscope for measuring the lateral Casimir force between corrugated
surfaces (see the text for further discussion).
\label{fig2}}
\end{figure}
The schematic of the experimental setup is shown in Figure~2 \cite{43}.
A sinusoidally corrugated grating was vertically mounted on the piezoelectric tube.
A smooth sphere was placed at the end of the cantilever of an atomic force microscope.
A mica sheet was attached to the bottom of this sphere. The second sphere was attached
to the bottom free end of the mica sheet. Then, the sinusoidal corrugations were
imprinted on the second sphere from the grating as a template using a special
procedure \cite{43}. This ensures a parallelism of the corrugation axes on both
test bodies. The interacting surfaces were covered with an Au layer.
\begin{figure}[t]
\centering
\vspace*{-5.cm}
\includegraphics[width=18 cm]{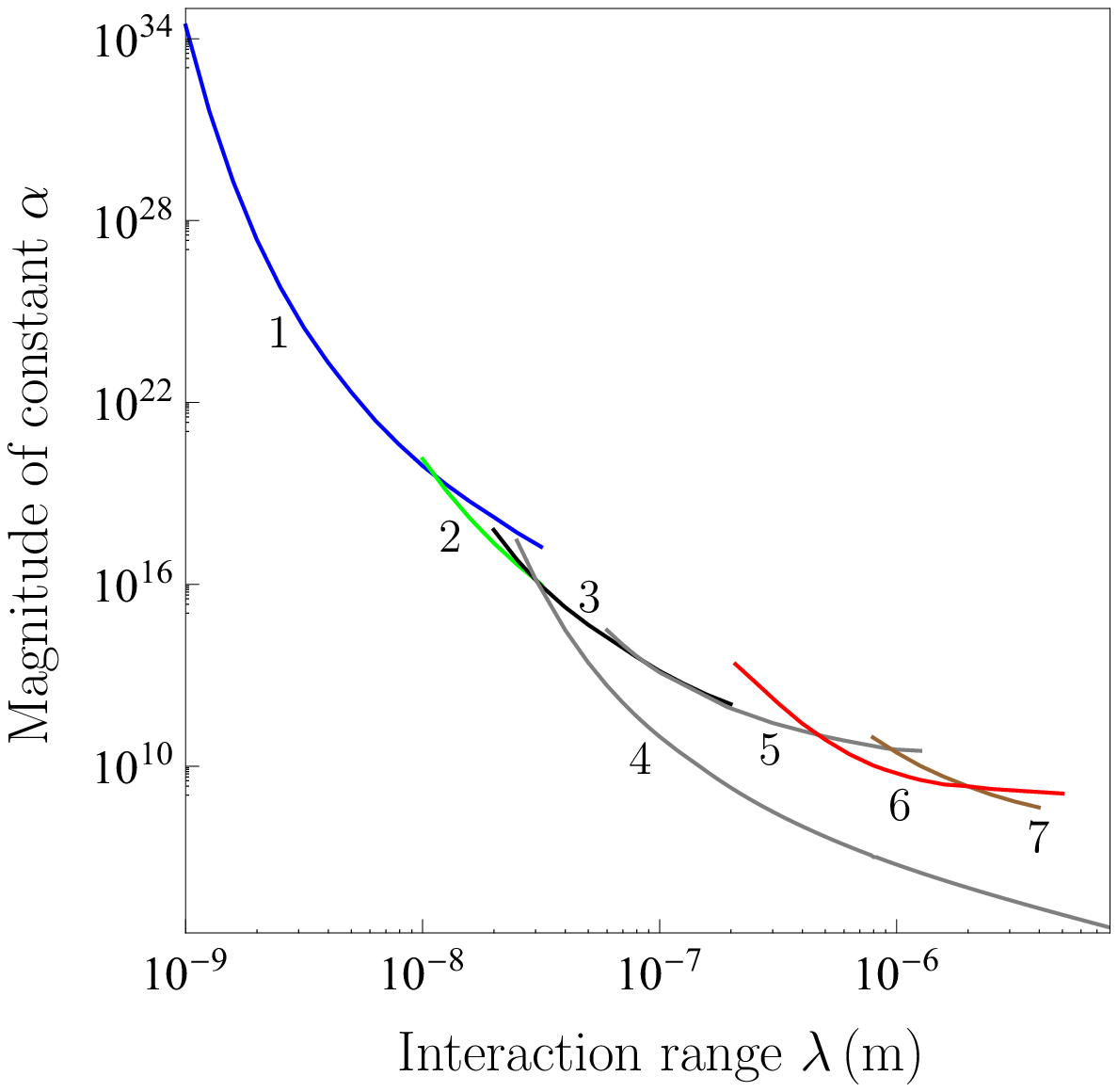}
\vspace*{-10.5cm}
\caption{The constraints on Yukawa-type correction to Newton's gravitational
law following from measuring the lateral and normal Casimir forces between
corrugated surfaces (lines 1 and 2, respectively), the effective Casimir
pressure (line 3), from the second and first Casimir-less experiments
(lines 4 and 5, respectively), from sensing the difference of lateral
forces (line 6), and from the torsion-pendulum experiment (line 7) are
shown as functions of the interaction range.
The regions above each line are excluded and below each line
are allowed.\label{fig3}}
\end{figure}

The lateral Casimir force $F_{C,\,\rm lat}$ was measured at different values of the
phase shift $\Phi_i$ at different separations from 120 to 190~nm, and the total
experimental errors $\Delta_i F_{C,\,\rm lat}$ have been determined at the
95\% confidence level.
The lateral Casimir force between the sinusoidally corrugated surfaces was also
calculated using the generalization of the Lifshitz theory based on the Rayleigh
scattering approach, and good agreement with the measurement data in the limits
of the errors $\Delta_i F_{C,\,\rm lat}$ was observed. This means that the
lateral force of Yukawa-type, which also should arise due to the presence of
corrugations, satisfies the inequality
\begin{equation}
| F_{\,\rm Yu,lat}(a_i,\Phi_i)|\leqslant \Delta_i F_{C,\,\rm lat}.
\label{eq5}
\end{equation}

The hypothetical force $F_{\,\rm Yu,lat}$ in the experimental configuration of
Figure~2 was calculated in \cite{45} using (\ref{eq1}) and
(\ref{eq2}), where a dependence on $\Phi$ arises for the corrugated
plates, and subsequent negative differentiation with respect to $\Phi$
 similar to (\ref{eq4}).
This gave the possibility to determine the value of $\alpha$ and $\lambda$
which satisfy (\ref{eq5}).
The obtained results are shown in Figure~3 by the line 1, where here and below
the region of the $(\lambda,\alpha)$ plane above the line is excluded and the
region below it is allowed by the experimental results. The constraints of the
line 1 are stronger by up to a factor of $2.4\times 10^7$ than the previously
known constraints obtained at a high confidence level \cite{45}
(although the strength of constraints obtained in \cite{46} from the
measurement data of \cite{47} is slightly stronger,  their confidence level
remains indefinite \cite{17,48}).

At larger interaction range $\lambda>11.6~$nm the stronger constraints on the
strength of Yukawa interaction follow from measurements of the normal Casimir
force between sinusoidally corrugated surfaces of a sphere and a plate at
different orientation angles of corrugations \cite{49,50}.
The schematic of the experimental setup is shown in Figure~4 \cite{49}.
In this case, unlike Figure~2, a sinusoidally corrugated grating was mounted on the
piezoelectric tube laterally. An initially smooth sphere was attached to the end of
the cantilever of an atomic force microscope. The sinusoidal corrugations on it were
imprinted from the grating as a template. Both interacting surfaces were coated
with an Au layer.
\begin{figure}[t]
\centering
\vspace*{-4.5cm}
\includegraphics[width=16 cm]{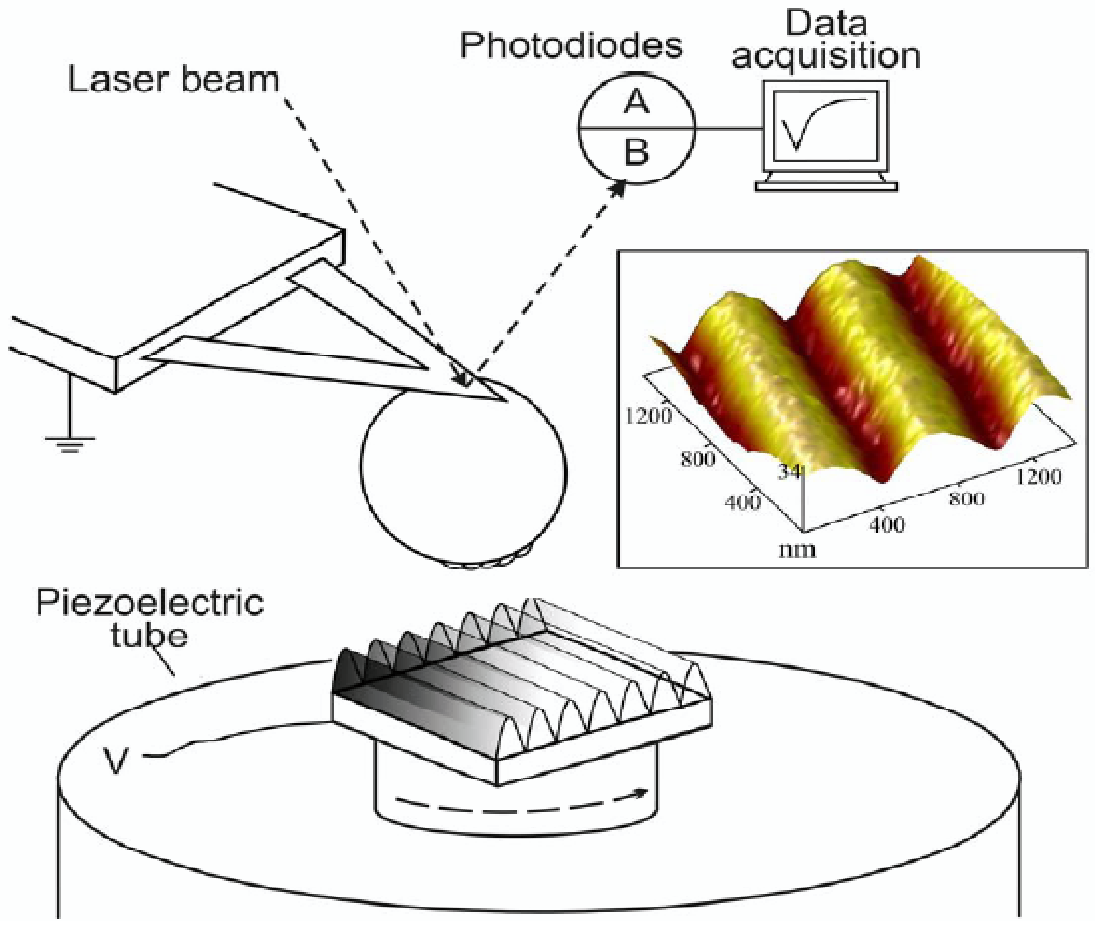}
\vspace*{-10cm}
\caption{Schematic diagram of the experimental setup using an atomic force
microscope for measuring the normal Casimir force between corrugated
surfaces (see the text for further discussion).
\label{fig4}}
\end{figure}

The normal Casimir force $F_{C,\,\rm nor}$ was measured at different separations from
127 to 200~nm for different angles $\theta_i$ between the axes of corrugations from
0 to $2.4^{\circ}$ at $T=300~$K. The total experimental errors
$\Delta_i F_{C,\,\rm nor}$ have been determined at the 67\% confidence level.
However, we have recalculated them to the 95\% confidence level in order to obtain
constraints consistent with other results discussed here. In the limits of these
errors the measurement data were found in good agreement with the theoretical
Casimir force calculated at $T=300~$K using the method of derivative expansion
\cite{51,52}. Thus, the perpendicular to the surface Yukawa force should
satisfy the inequality
\begin{equation}
| F_{\,\rm Yu,nor}(a_i,\Phi_i)|\leqslant \Delta_i F_{C,\,\rm nor}.
\label{eq6}
\end{equation}

The Yukawa force in the configuration of Figure~4 was calculated in \cite{53}.
The resulting constraints on $\alpha$ and $\lambda$ are shown by the line 2 in
Figure~3. They were considered as the strongest ones up to 17.2~nm \cite{54}.
\begin{figure}[!b]
\centering
\vspace*{-5.5cm}
\hspace*{-2cm}
\includegraphics[width=18 cm]{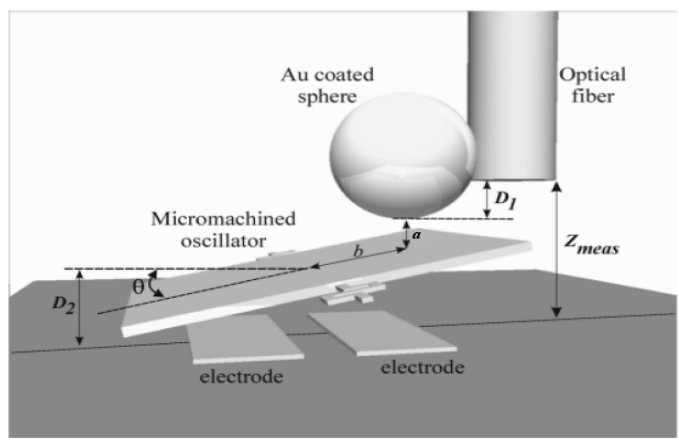}
\vspace*{-13.5cm}
\caption{Schematic diagram of the experimental setup using a micromechanical
oscillator for determination of the effective Casimir pressure between two
parallel plates (see the text for further discussion).
\label{fig5}}
\end{figure}

At $\lambda>17.2~$nm the stronger constraints have been obtained from
measuring the effective Casimir pressure between two parallel plates by means of
a micromechanical torsional oscillator \cite{55,56}.
The schematic diagram of this measurement is shown in Figure~5 \cite{17}.
The oscillator consisted of an Au-coated plate  suspended at two opposite points
on the midplane. Under the plate there were two electrically contacted electrodes
used to measure the capacitance and to induce oscillation in the plate at the
resonant frequency of the oscillator. Above the plate there was an Au-coated
sphere of radius $R$ attached to the side of an Au-coated optical fiber.
The separation between the plate and the sphere was determined as
\begin{equation}
a=z_{\rm meas}-D_1-D_2-b\theta,
\label{eq7}
\end{equation}
\noindent
where the meanings of $z_{\rm meas}$, $D_1$, $D_2$, $\theta$, and $b$ are clear
from Figure~5.
In the dynamic regime used in this experiment, an immediately measured quantity
was a shift of the resonant frequency of the oscillator under an impact of the
Casimir force $F_{C,sp}$ acting between a sphere and a plate. The solution for
the equation of motion of this oscillator allows recalculation of the
frequency shift into the gradient of the Casimir force
$\partial F_{C,sp}/\partial a$. Using the proximity force approximation \cite{17},
which is valid in this case up to a fraction of a percent \cite{57,58}, one
obtains the effective Casimir pressure between two parallel plates
\begin{equation}
P_C(a,T)=-\frac{1}{2\pi R}\,\frac{\partial F_{C,sp}(a,T)}{\partial a}.
\label{eq8}
\end{equation}

The effective Casimir pressure (\ref{eq8}) was measured within the separation
range from 162 to 746~nm at $T=300~$K, and the total experimental errors $\Delta_i P_C$
were determined at the 95\% confidence level. A very good agreement was found between
the measurement data and theoretical predictions of the Lifshitz theory using the
optical data for the complex index of refraction of Au extrapolated down to zero
frequency by the dissipationless plasma model \cite{55,56}. An alternative extrapolation
using the dissipative Drude model was excluded by the data. A discussion of both
extrapolations can be found in \cite{17,48,59}. Later an exclusion of the Drude
extrapolation was conclusively confirmed by the experiment \cite{60} where the
theoretical predictions from two alternative extrapolations differ by up to a factor
of 1000.
{  We emphasize that a difference between predictions of the excluded and
confirmed versions of the Lifshitz theory cannot be modeled by the Yukawa force
\cite{60a}.}
This means that the effective Yukawa pressure between two parallel plates
should satisfy the inequality
\begin{equation}
|P_{\,\rm Yu}(a_i)|\leqslant\Delta_iP_C.
\label{eq9}
\end{equation}

The Yukawa-type pressure between two parallel Au-coated plates was calculated in
\cite{55,56} and used for constraining the values of $\alpha$ and $\lambda$ in accordance
to (\ref{eq9}). The obtained constraints are shown by line 3 in Figure~3.
For some time these constraints were considered as the strongest ones up to $\lambda=89~$nm.

At larger $\lambda$ the strong constraints on $\alpha$ were obtained from the so-called
{\it Casimir-less} experiment where an impact of the Casimir force was nullified \cite{61}.
This experiment was performed using the micromechanical oscillator shown in Figure~5.
The plate of this oscillator was, however, composite. It consisted of two halves
coated by Au and Ge films with a common Au overlayer. The oscillator was moved back and
forth below the sphere, so that only a difference of forces from two halves of the plate
has been measured. In doing so a common Au overlayer was sufficiently thick to make the
Casimir forces from two halves of the plate equal. As to the Yukawa-type forces between
the sphere and the plate halves coated with Au and Ge films, they were
{  predicted to be} different.
The sensitivities $\Delta_i F_{\rm diff}$ of this setup to the differential force determined
at the 95\% confidence level were of the order of 1~fN at separations $a_i$ varying
from 150 to 500~nm. This means that the difference of Yukawa forces from the Au and Ge
films should satisfy the inequality
\begin{equation}
|F_{\,\rm Yu,Au}(a_i)-F_{\,\rm Yu,Ge}(a_i)|\leqslant\Delta_iF_{\rm diff}.
\label{eq10}
\end{equation}

Both these Yukawa forces were calculated and used for constraining of $\alpha$ and
$\lambda$ in \cite{61}. The constraints following from (\ref{eq10}) are shown by line 5 in
Figure~3 (line 4 is discussed in the end of this section). They were the best ones
within  the range of $\lambda$ from 89 to 465~nm.

At larger $\lambda$ stronger constraints were obtained by sensing the difference of lateral
forces between an Au sphere attached to the end of soft cantilever and a density modulation
source mass \cite{62}. To make a cantilever sensitive to the lateral force, it was placed
normal to the source-mass surface. The obtained results are shown by line 6 in Figure~3.
At $\lambda=2.09~\mu$m they are replaced by the stronger ones found from measurement of
the Casimir force between the Au-coated surfaces of a plate and a spherical lens of
centimeter-size radius by means of the torsion pendulum \cite{63}. These constraints
are shown by line 7 in Figure~3. They extend from $\lambda=2.09~\mu$m to
$\lambda=3.16~\mu$m. For larger $\lambda$ the strongest constraints on the Yukawa-type
interaction follow from the gravitational experiments (see Section~4).

In the end of this section, we present the results of the second Casimir-less experiment
\cite{26} which improved the results of the first one \cite{61} significantly and
resulted in stronger constraints over a wider interaction range than those ones obtained
from several experiments listed above. As compared to the first Casimir-less experiment,
the experimental setup for measuring the difference of forces was considerably upgraded.
A differential force was measured between an Au-coated sphere and either an Au or a Si
sectors of the rotating disc coated with an Au overlayer of sufficient thickness.
In this way, the difference of Casimir forces between the sphere and the two
neighboring sectors was again nullified, and the measurement result was determined
by only the difference of Yukawa forces between the same sphere and either an Au or a Si
sectors. Furthermore, the force sensitivity of the setup was improved, so that the
minimum detectable force $\Delta_i F_{\rm diff}$ varied from 0.1 to 0.2~fN when
separation distance increased from 200 to 1000~nm.

The constraints on the Yukawa-type interaction have been again derived from inequality
(\ref{eq10}) where the lower index Ge should be replaced with Si. The obtained results
\cite{26} are shown by line 4 in Figure~3. It is seen that they extend the previously
known constraints over a wide interaction range and strengthen them by up to a factor
of $10^3$. The constraints of line 4 are the strongest ones in the interaction
region from $\lambda=40~$nm to $\lambda=8~\mu$m. Thus, for $\lambda>40~$nm they
improve the constraints following from measurements of the Casimir pressure and all
other constraints obtained from measuring the Casimir force at larger $\lambda$.

A summary of the strongest constraints on the Yukawa-type interaction over the
interaction range from 1~nm to 1~mm obtained from measurements of the Casimir force
and other laboratory experiments is presented in Section~4.

\section{Constraints on the Coupling of Axions to Nucleons
from the Casimir Effect}

As was mentioned in Section~1, the effective potential due to an exchange of one
axion between two nucleons depends on their spins $\mbox{\boldmath$\sigma$}_1$ and
$\mbox{\boldmath$\sigma$}_2$. An explicit form of this potential is given by
\cite{14,15,64}
\begin{equation}
V_{an}^{(1)}(\mbox{\boldmath$r$}_{12};\mbox{\boldmath$\sigma$}_1,\mbox{\boldmath$\sigma$}_2)
=\frac{g_{an}^2}{16\pi m^2}\left[(\mbox{\boldmath$\sigma$}_1,\mbox{\boldmath$k$})
(\mbox{\boldmath$\sigma$}_2,\mbox{\boldmath$k$})
\left(\frac{m_a^2}{r_{12}}+3\frac{m_a}{r_{12}^2}+\frac{3}{r_{12}^3}\right)
-(\mbox{\boldmath$\sigma$}_1,\mbox{\boldmath$\sigma$}_2)
\left(\frac{m_a}{r_{12}^2}+\frac{1}{r_{12}^3}\right)\right]\,
e^{-m_ar_{12}}.
\label{eq10a}
\end{equation}
\noindent
Here, we assume that the dimensionless coupling constants of an axion to a proton and
a neutron are approximately equal $g_{an}=g_{ap}$. The nucleon mass is notated $m$, the
axion mass is $m_a$, and
$\mbox{\boldmath$k$}=\mbox{\boldmath$r$}_{12}/|\mbox{\boldmath$r$}_{12}|$ is the unit
vector. Equation (\ref{eq10a}) is valid for both the QCD and GUT axions.

The effective potential (\ref{eq10a}) averages to zero after an integration over the
volumes of two macrobodies. Because of this, the process of one-axion exchange does not
lead to any additional force which could be searched for in the Casimir experiments
(an experiment on measuring the Casimir force between two polarized test bodies has
been proposed \cite{65} but is not realized yet). If, however, the axion-to-nucleon
interaction is described by the pseudoscalar Lagrangian characteristic for GUT axions,
a simultaneous exchange by two axions results in the spin-independent effective
potential \cite{14,15,66}
\begin{equation}
V_{an}^{(2)}(r_{12})=-\frac{g_{an}^4}{32\pi^3 m^2}
\frac{m_a}{r_{12}^2}K_1(2m_ar_{12}),
\label{eq11}
\end{equation}
\noindent
where $K_1(z)$ is the modified Bessel function of the second kind.
This result, which is valid under the condition $r_{12}\gg m^{-1}$, resembles the
Yukawa-type potential (\ref{eq1}). As to the pseudovector Lagrangian, which can be also
used to describe the interaction of axions to nucleons, it is characteristic for the QCD
axion models. For this Lagrangian, the form of effective potential describing the
two-axion exchange remains unknown \cite{16}. The results presented below in this
section are based on the use of (\ref{eq11}) and, thus, are applicable to the GUT
axions and axionlike particles.

Similar to the case of Yukawa-type forces, the interaction energy of two test bodies
$V_1$ and $V_2$ due to an exchange of two GUT axions between their protons and neutrons
is given by
\begin{equation}
E_{an}^{(2)}(a)=-\frac{g_{an}^4m_a}{32\pi^3m^2}n_1n_2\int_{V_1}\int_{V_2}
d\mbox{\boldmath$r$}_1d\mbox{\boldmath$r$}_2
\frac{K_1(2m_ar_{12})}{r_{12}^2},
\label{eq12}
\end{equation}
\noindent
where $n_1$ and $n_2$  are the numbers of nucleons per unit volume of the first and
second test bodies, respectively. Then, the additional force arising due to two-axion
exchange is given by
\begin{equation}
F_{an}^{(2)}(a)=-\frac{\partial E_{an}^{(2)}(a)}{\partial a}.
\label{eq13}
\end{equation}

Different experiments on measuring the Casimir force were used to constrain the
axion-to-nucleon interaction in the same way as described in Section~2 for the
interaction of Yukawa type. We begin with the second Casimir-less experiment \cite{26}
(see Section~2). Its results were used for constraining axions in \cite{37}.
For this purpose, the difference of additional forces arising in the experimental
configuration due to two-axion exchange was found using (\ref{eq12}) and (\ref{eq13}).
Taking into account that no force was registered up to the sevsitivities
$\Delta_i F_{\rm diff}$ varying from 0.1 to 0.2~fN, this difference should satisfy
the inequality
\begin{equation}
|F_{an,\rm Au}^{(2)}(a_i)-F_{an,\rm Si}^{(2)}(a_i)|\leqslant\Delta_i F_{\rm diff}.
\label{eq14}
\end{equation}

The constraints on an axion-to-nucleon interaction are usually displayed graphically as
$g_{an}^2/(4\pi)$ depending on the axion mass $m_a$. Similar to the Yukawa-type
interaction, the region of the $[m_a,g_{an}^2/(4\pi)]$ plane above the line is excluded
by the results of respective experiment and below the line is allowed. By the
line 1 in Figure~6 we show the constraints on the axion mass and interaction constant
following from the Casimir-less experiment. Below they are compared with similar
constraints following from several other experiments.
\begin{figure}[!b]
\centering
\vspace*{-5.cm}
\includegraphics[width=18 cm]{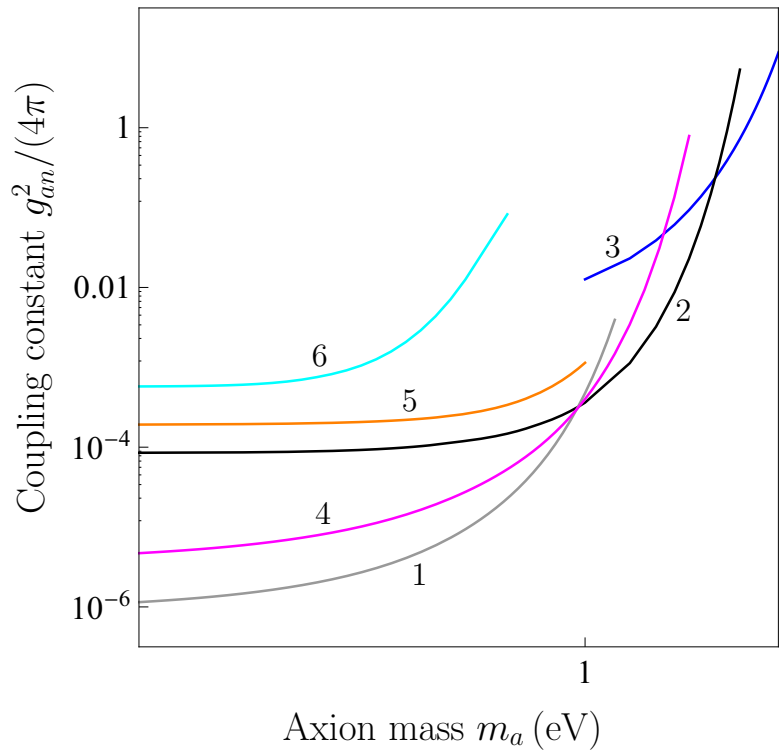}
\vspace*{-10.5cm}
\caption{The constraints on the coupling constant of axion to nucleon
following from the Casimir-less experiment, experiments on measuring the
Casimir pressure and lateral Casimir force (lines 1, 2, and 3,
respectively) and from experiments on measuring the difference of Casimir
forces, gradient of the Casimir force and Casimir-Polder force
(lines 4, 5, and 6, respectively) are shown as functions of the axion mass.
The regions above each line are excluded and below each line
are allowed.
\label{fig6}}
\end{figure}

Next, we consider an experiment on measuring the effective Casimir pressure by means
of a micromechanical oscillator shown schematically in Figure~5 \cite{55,56}.
This experiment was used for constraining the parameters of axions in \cite{67}.
For this purpose, an additional pressure $P_{an}^{(2)}$ was calculated arising
in the experimental configuration due to the two-axion exchange. The constraints were
obtained from the inequality
\begin{equation}
|P_{an}^{(2)}(a_i)|\leqslant \Delta_i P_C\, ,
\label{eq15}
\end{equation}
\noindent
where the meaning of $\Delta_i P_C$ is explained in Section~2.
These constraints are shown by the  line 2 in Figure~6. As is seen in Figure~6,
the constraints on an axion following from measuring the Casimir pressure become
stronger that those found from the Casimir-less experiment for axion masses
$m_a>0.9~$eV.

For axion masses $m_a>8~$eV the stronger constraints on $g_{an}$ follow from measurements
of the lateral Casimir force between corrugated surfaces (see Figure~2 for a schematic of
the experimental setup and \cite{43,44} for a description of the measurement procedure).
To obtain these constraints, the additional force $F_{an,\rm lat}^{(2)}$ was calculated
in \cite{68}. Then the constraints were obtained from the inequality
\begin{equation}
|F_{an,\rm lat}^{(2)}(a_i,\Phi_i)|\leqslant \Delta_i F_{C,\rm lat}\, ,
\label{eq16}
\end{equation}
\noindent
where  $\Delta_i F_{C,\rm lat}$ is the total experimental error in measuring the
lateral Casimir force (see Section~2  for details). These constraints are shown by
the line 3 in Figure~6.

In addition to the constraints shown by the lines 1--3, which are the strongest
ones over the respective regions of axion masses, in Figure~3 we also present the
constraints obtained from three more Casimir experiments. Although these experiments
lead to somewhat weaker constraints, they are useful by providing a qualitatively
similar picture based on the independent measurement data.

The line 4 in Figure~6 was obtained in \cite{69} from the measure of agreement
between experiment and theory in measuring the difference in Casimir forces between
a Ni-coated sphere and either an Au or a Ni sectors of a rotating disc coated with
a thin Au overlayer \cite{60}. As opposed to the Casimir-less experiments \cite{26,61},
the Au overlayer was sufficiently thin in this case and did not make equal the
Casimir forces when a sphere was spaced above an Au and a Ni sectors. It is seen
that the line 4 is sandwiched between the lines 1 and 2 discussed above.
Thus, the constraints following from measuring the difference in Casimir forces are
in agreement with those obtained from the Casimir-less experiment and experiment on
measuring the Casimir pressure.

The line 5 in Figure~6 was obtained in \cite{70} from the dynamic experiment
measuring the gradient of the Casimir force between an Au-coated sphere and an
Au-coated plate performed by means of the dynamic force microscope \cite{71}.
Finally, the line 6 in Figure~6 follows \cite{72} from the experiment on
measuring the Casimir-Polder force between ${}^{87}$Rb ultracold atoms belonging
to the Bose-Einstein condensate and a silica glass plate \cite{73}.
As is seen in Figure~6, the lines 5 and 6 indicate the weaker constraints
on the coupling constant of axions to nucleons than those shown by the line 2
following from measuring the effective Casimir pressure.

In the next section the constraints of Figure~6 are compared with some other
constraints obtained from the laboratory experiments unrelated to measurements
of the Casimir force.

\section{Constraints from Measuring the Casimir Force in Nanometer Separation Range
and Other Laboratory Experiments}

Here, we consider the constraints on the Yukawa-type hypothetical interaction and on
the coupling constant of axions to nucleons obtained recently \cite{74} from
measuring the Casimir force between an Au-coated sphere and a silicon carbide
(SiC) plate
at $T=300~$K in N${}_2$ atmosphere \cite{75}. This experiment has the advantage that
the surfaces of both interacting bodies are very smooth. This allowed to measure the
Casimir force by means of an atomic force microscope over the separation region from
10 to 200~nm. The experimental results for the Casimir force were compared with
theoretical predictions of the Lifshitz theory. For this purpose, the optical
properties of SiC plate have been determined by means of ellipsometers \cite{76},
as well as the optical properties of Au coating on the sphere which were characterized
earlier \cite{77}. Both the experimental and theoretical errors were determined
at the 67\% confidence level. In \cite{74} the total measure of agreement between
experiment and theory $\Delta_i F_C$ was found at the 95\% confidence level.
Within these errors no additional force was registered  originating  either from
the non-Newtonian gravity or from the axion-to-nucleon interaction.

We begin with the Yukawa-type interaction (\ref{eq1}). From (\ref{eq1})--(\ref{eq3})
one obtains the following expression for the Yukawa force between an Au-coated sphere
and a plate \cite{74}:
\begin{equation}
F_{\,\rm Yu}(a)=-4\pi^2G\alpha\rho_1\rho_2\lambda^3e^{-a/\lambda}\left[
\Psi(R,\lambda)-\Psi(R-d,\lambda)\,e^{-d/\lambda}\right],
\label{eq17}
\end{equation}
\noindent
where $\rho_1$ and $\rho_2$ are the densities of Au and SiC, respectively,
$d$ is the thickness of an Au coating on the sphere, and the function $\Psi$ is
given by
\begin{equation}
\Psi(r,\lambda)=r-\lambda+(r+\lambda)\,e^{-2r/\lambda}.
\label{eq18}
\end{equation}
\noindent
In the derivation of (\ref{eq17}) it was taken into account that the sphere core
material (borosilicate glass) gives only a minor contribution to the Yukawa force
which does not influence the obtained results.

The constraints on the Yukawa-type correction to Newtonian gravity have been found
from the inequality
\begin{equation}
|F_{\,\rm Yu}(a_i)|\leqslant \Delta_i F_C.
\label{eq19}
\end{equation}
\noindent
They are shown by the line labeled "new" in Figure~7. In the same figure we reproduce from
Figure~3 the lines 1, 2, 3, and 4 obtained from measuring the lateral and normal
Casimir forces between corrugated surfaces, the Casimir pressure, and from the
second Casimir-less experiment, respectively.
As is seen in Figure~7, the new constraints are
stronger than those of line 1 within the interaction range from 1 to 3.7~nm.
The largest strengthening by the factor of $5\times 10^5$ is reached at
$\lambda=1~$nm.
\begin{figure}[t]
\centering
\vspace*{-8.cm}
\includegraphics[width=18 cm]{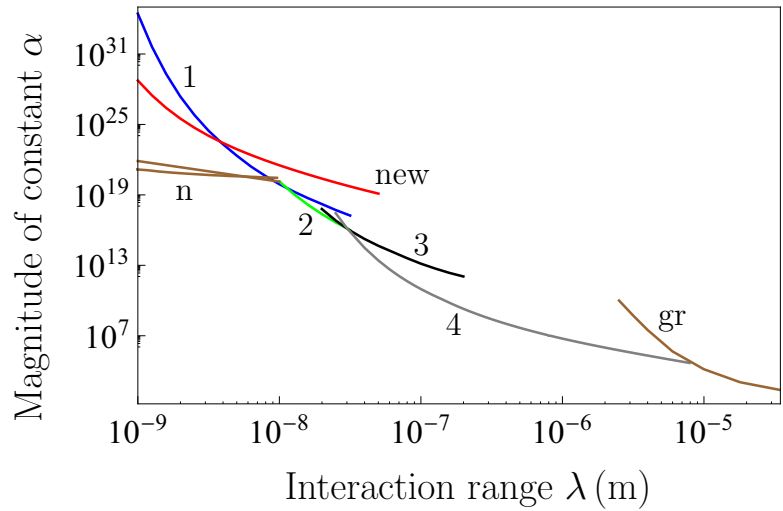}
\vspace*{-10.5cm}
\caption{The constraints on Yukawa-type correction to Newton's gravitational
law following from recent measuring the Casimir force at nanometer separations
(line labeled "new"), from experiments on neutron scattering (lines labeled
"n"), and from the Cavendish-type experiment (line labeled "gr")
are shown as functions of the interaction range.
The lines 1--4, which also indicate the strongest constraints within the
respective interaction ranges, are reproduced from Figure~3.
The regions above each line are excluded and below each line
are allowed.
\label{fig7}}
\end{figure}

Now we compare the constraints on the Yukawa interactions  obtained from the Casimir
effect and from some other laboratory experiments. The strongest constraints on $\alpha$
in the nanometer interaction range have been found from the experiments on neutron
scattering \cite{23,24}. They are indicated in Figure~7 by the two lines labeled
"n". As is seen in Figure~7, the neutron scattering results in the strongest
constraints on $\alpha$ up to $\lambda=10~$nm.

Over a wide interaction range from  $\lambda=10~$nm to $\lambda=8~\mu$m the
strongest constraints on $\alpha$ follow at the moment from the Casimir physics.
In doing so measurements of the lateral and normal Casimir forces
between corrugated surfaces lead to the most strong results only within the
very narrow ranges from 10 to 11,6~nm and from 11.6 to 17.2~nm, respectively
(lines 1 and 2 in Figure~7). The measurement of the Casimir pressure results in the
strongest constraints in the range from   $\lambda=17.2~$nm to  $\lambda=40~$nm
(see line 3 in Figure~7). Finally, the second Casimir-less experiment leads to
the most strong constraints in the wide range from  $\lambda=40~$nm to
$\lambda=8~\mu$m (line 4 in Figure~7).

For larger $\lambda$ the strongest constraints on the Yukawa-type corrections to
Newtonian gravitational law are currently obtained from the gravitational
experiments. Specifically, within the interaction range from $\lambda=8~\mu$m to
$\lambda=1~$cm the strongest constraints follow from the Cavendish-type
experiments \cite{18,19,20}. Thus, in the range from 8 to $9~\mu$m they were obtained
in \cite{18}, in the range from $9~\mu$m to 4~mm in \cite{19}, and for larger
$\lambda$ up to 1~cm in \cite{20}. The initial part of the line representing these
constraints in the region from 8 to $34~\mu$m is labeled "gr" in Figure~7.
Recently the constraints of \cite{19} were strengthened by up to a factor of 3
in the region of $\lambda$ from 40 to $350~\mu$m by means of the improved
Cavendish-type experiment \cite{78}.

Now we deal with the constraints on an axion which follow \cite{74} from measuring
the Casimir force in nanometer separation region \cite{75}. An expression for the
additional force acting between an Au-coated sphere and a SiC plate due to
two-axion exchange is obtained from (\ref{eq12}) and (\ref{eq13}) \cite{74}
\begin{equation}
F_{an}^{(2)}(a)=-\frac{n_1n_2g_{an}^4}{32\pi m_am^2}\int_1^{\infty}\!\!\!du
\frac{\sqrt{1-u^2}}{u^3}\,e^{-2m_a au}\left[
\chi(R,m_a u)-\chi(R-d,m_au)e^{-2m_a d u}\right],
\label{eq20}
\end{equation}
\noindent
where the function $\chi$ is defined as
\begin{equation}
\chi(r,z)=r-\frac{1}{2z}+\left(r+\frac{1}{2z}\right)\,e^{-2rz}
\label{eq21}
\end{equation}
\noindent
and $n_1,\,\,n_2$ are the numbers of nucleons per unit volume of Au and
SiC \cite{4}.

The constraints on the coupling constant of axions to nucleons were obtained
from the inequality
\begin{equation}
|F_{an}^{(2)}(a_i)|\leqslant \Delta_i F_C,
\label{eq22}
\end{equation}
\noindent
which is similar to (\ref{eq19}) used to constrain the Yukawa-type interactions
from the results of the same experiment. This constraints are shown by line
"new" in Figure~8. In this figure the solid lines 1, 2, and 3 following from the
second Casimir-less experiment, measurements of the Casimir pressure, and the
lateral Casimir force between corrugated surfaces, respectively, are reproduced
from Figure~6. As is seen in Figure~8, the new constraints are stronger than
those found from measuring the lateral Casimir force for axion masses
$m_a>17.8~$eV.
\begin{figure}[t]
\centering
\vspace*{-7.5cm}
\includegraphics[width=18 cm]{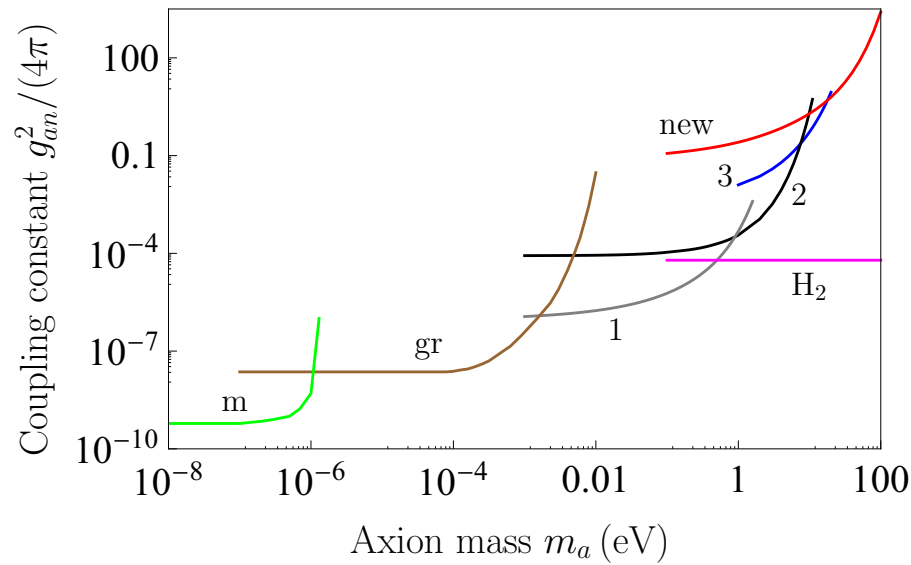}
\vspace*{-10.5cm}
\caption{The constraints on the coupling constant of axion to nucleon
following from recent measuring the Casimir force at nanometer separations
(line labeled "new"), from the magnetometer measurements (line labeled
"m"), from the Cavendish-type experiment (line labeled "gr"), and from
the experiment using beams of molecular hydrogen (line labeled "H${}_2$")
are shown as functions of the axion mass.
The lines 1--3, where line 1 indicates the strongest constraints within
some range of axion masses, are reproduced from Figure~6.
The regions above each line are excluded and below each line
are allowed.
\label{fig8}}
\end{figure}

We are coming now to a comparison of the above constraints on an axion-to-nucleon
coupling constant obtained from the Casimir effect with those following from some
other laboratory experiments. Strong constraints on the coupling constant of axions
to neutrons over the wide range of axion masses from $m_a=10^{-4}~\mu$eV to
$m_a=1~\mu$eV were obtained from the magnetometer measurements employing
spin-polarized K and ${}^3$He atoms \cite{35}. These constraints in the range of
$m_a$ from $10^{-2}~\mu$eV to $1~\mu$eV are shown by line labeled "m" in
Figure~8. We note that the constraints found by means of a magnetometer are derived
from the one-axion exchange between two neutrons which is described by the
effective potential (\ref{eq10a}). Because of this they are equally applicable
to both the QCD and GUT axions.

Within the wide range of $m_a$ from $1~\mu$eV to 1.7~meV the strongest constraints
on the axion-to-nucleon coupling constant valid for the GUT axions follow from
the gravitational experiments of Cavendish type \cite{19}. These constraints
derived in \cite{36} are shown by line labeled "gr" in Figure~8.
For $m_a>1.7~$meV the strongest constraints on $g_{an}$ are given by line 1
reproduced from Figure~6 which was obtained from the Casimir-less experiment.

In the case of axions, this experiment leads to competitive constraints up to
$m_a=0.5~$eV. For $m_a>0.5~$eV the constraints following from the Casimir physics
(lines 1--3 and "new" in Figure~8) become much weaker than the constraints
obtained from experiment on measuring the forces between protons in the beam of
molecular hydrogen \cite{38,39}. These constraints are shown by line labeled
"H${}_2$" in Figure~8. For axions with  $m_a>200~$eV the constraints of line
"H${}_2$" were further strengthened by comparing with theory the experimental
results on the nuclear magnetic resonance for nucleons in deuterated molecular
hydrogen \cite{39}.

\section{Proposed Experiments}
In this section we consider several experiments suggested in the literature which
allow further strengthening the constraints on non-Newtonian gravity and axionlike
particles but are not performed yet.

All the experiments on measuring the Casimir interaction considered above have been
planned and performed for the purposes which were not directly connected with
investigation of hypothetical interactions. The obtained constraints can be
considered as some by-product. In \cite{79} the parameters of these experiments were
optimized in order the stronger constraints  could be obtained on both the
Yukawa-type interaction and the interaction due to two-axion exchange. It was shown
\cite{79} that the prospective constraints are up to an order of magnitude stronger
than those ones already obtained and outlined above.

All precise measurements of the Casimir interaction performed during the last few
years employed the sphere-plate geometry and the separation distances below a
micrometer. For this reason much attention was attracted to the proposed CANNEX
(Casimir And Non-Newtonian force EXperiment) test designed to measure the Casimir
force between two parallel plates spaced at separations up to a few micrometers
\cite{80,81}. In \cite{82} the parameters of  this experiment were optimized
for obtaining  the strongest constraints on both the non-Newtonian gravity and
axionlike particles. Here, we illustrate the potentialities of the proposed setup.

In Figure~9, the constraints on the Yukawa interaction constant $\alpha$, which could
be obtained from the CANNEX test, are shown by the line labeled "proposed".
For comparison purposes, the lines 4--7 and "gr" following from experiments discussed in
Sections~2 and 4 are reproduced from Figure~7. As is seen in Figure~9, the proposed line
provides an order of magnitude stronger constraints on $\alpha$ than those given by
line 4 following from the Casimir-less experiment. It is seen also that the proposed
constraints encroach on the province where the Cavendish-type experiments presently
lead to the strongest constraints shown by the line labeled "gr".
\begin{figure}[!b]
\centering
\vspace*{-7.5cm}
\includegraphics[width=18 cm]{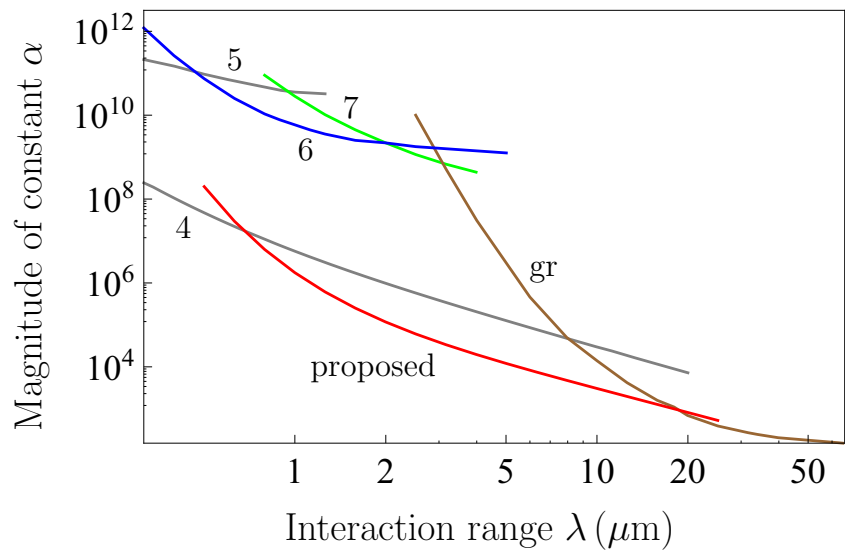}
\vspace*{-10.5cm}
\caption{The constraints on Yukawa-type correction to Newton's gravitational
law which can be obtained from the proposed CANNEX test are shown by the line
"proposed" as a function of the interaction range. For comparison purposes,
the lines 4--7 and the line labeled "gr" are reproduced from Figures~3 and 7,
respectively. The regions above each line are excluded and below each line
are allowed. \label{fig9}}
\end{figure}

A similar state of affairs holds for the additional interaction between two parallel
plates due to two-axion exchange. The constraints on $g_{an}$, which could
be obtained from the CANNEX test, are shown by the line labeled "proposed" in
Figure~10. For comparison purposes, the lines 1, 2, and labeled "gr" are reproduced
from Figure~8. From Figure~10 it is seen that the proposed line suggests stronger
constraints than those indicated by line 1 following from the Casimir-less experiment.
Some additional potentialities are also suggested by the CANNEX setup in out of
thermal equilibrium conditions \cite{83}.
\begin{figure}[t]
\centering
\vspace*{-7.5cm}
\includegraphics[width=18 cm]{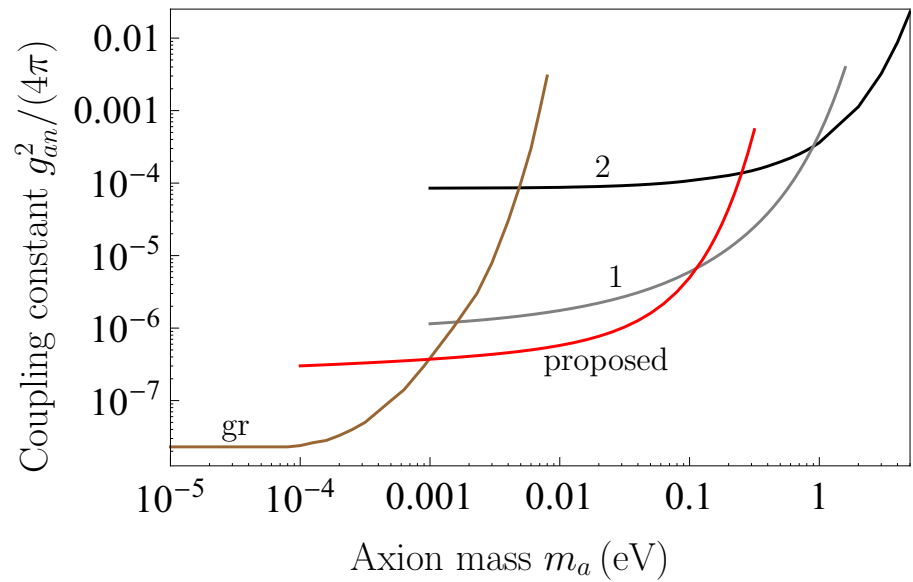}
\vspace*{-10.5cm}
\caption{The constraints on the coupling constant of axion to nucleon
which can be obtained from the proposed CANNEX test are shown by the line
"proposed" as a function of the axion mass.  For comparison purposes,
the lines 1, 2 and "gr" are reproduced from Figures~6 and 8,
respectively. The regions above each line are excluded and below each line
are allowed.\label{fig10}}
\end{figure}

There are several more proposed laboratory experiments aiming to strengthen the
constraints on non-Newtonian gravity and interaction of axionlike particles to
nucleons in the ranges of parameters considered above. Thus, it was suggested to
measure the Casimir-Polder force between a Rb atom and a movable Si plate separated
with a metallic shield in the form of an Au film \cite{84}. It is expected that
the experiment of this kind will help to obtain the stronger constraints on non-Newtonian
gravity in the interaction range of $\lambda\sim 1~\mu$m \cite{84}.
The respective strengthening of the constraints on $g_{an}$ is also expected.

According to \cite{85}, the weakly bound molecules can be used as sensitive sensors
for the non-Newtonian gravity. Calculations show that the precision spectroscopy of
these systems will allow to strengthen constraints on the interaction of Yukawa type
obtained from the experiments on neutron scattering (lines labeled "n" in
Figure~7) at least by the two orders of magnitude. Note that a few constraints
obtained from atomic spectroscopy in the range of extremely small $\lambda$ are
discussed in \cite{86}.

One more proposed detection of non-Newtonian gravity at short range exploits the
levitated sensors \cite{87}. It is based on the effect of normal mode splitting
in the optomechanical cavity and aims the detection of large extra dimensions
proposed in \cite{10,11} and the resulting Yukawa-type corrections to Newtonian
gravity \cite{12,13}.

In addition to the Yukawa-type interaction and forces arising due to two-axion
exchange between two nucleons, some other types of the fifth force are predicted
which are suppressed in the regions with high density of matter. These forces
are connected with such hypothetical scalar particles of variable mass as
chameleons and symmetrons. It was already pointed out \cite{79,80,81,88} that
experiments on measuring the Casimir force at large separations can be
used for a search of chameleons as possible candidates for the constituents
of dark energy. Recently the Casimir experiment was proposed \cite{89} which
should be capable to place stronger constraints on symmetrons.

\section{Discussion}

In this article, we have reviewed the present state of the art in constraining
the Yukawa-type corrections to Newton's law of gravitation and the
axion-to-nucleon interaction from laboratory experiments.
In was shown that measurements of the Casimir force and Casimir-less experiments
lead to strong constraints on the Yukawa interaction constant $\alpha$ and on
the coupling constant of axions to nucleons $g_{an}$ over the wide ranges of
$\lambda$ and $m_a$, respectively. The very recent results were presented on
constraining both $\alpha$ and $g_{an}$ from the experiment on measuring the
Casimir force between an Au-coated sphere and a SiC plate performed at short
separations down to 10~nm. These constraints are much stronger that those found
previously from measuring the lateral Casimir force between corrugated surfaces.

It should be noted, however, that within the interaction range of $\lambda$
below 10~nm and above $8~\mu$m the strongest constraints on the Yukawa-type
interaction follow not from the Casimir experiments, but from the laboratory
experiments of some other kind. Thus, for $\lambda<10~$nm the most strong
constraints on $\alpha$ were obtained from experiments on neutron scattering whereas
for $\lambda>8~\mu$m from the Cavendish-type experiments.
In a similar way, for axion masses $m_a$ below 1.7~meV the strongest laboratory
constraints on the coupling constant of axions to nucleons follow not from
the Casimir physics, but from the magnetometer measurements and experiments
of Cavendish type. At $m_a>0.5~$eV the strongest constraints on $g_{an}$
were obtained from measuring the forces between protons in the beam of
molecular hydrogen.

Thus, the future trends in this field of research are connected with
different experimental approaches to constraining the Yukawa-type forces
and the interaction of axions with nucleons.

\section{Conclusions}

To conclude, recent advances in experimental technique made possible performing
more precise laboratory experiments in different branches of physics.
The results of these experiments and the measure of their agreement with theory
were used for obtaining stringent laboratory constraints on the strength of
Yukawa-type interactions and the coupling constant of axions to nucleons.
In this article, we summarized the strongest constraints of this kind following
from the diversified experiments each of which was used for obtaining the best
result in some specific range of parameters. One may hope that by improving
the characteristics of these experiments and by implementing some of the recent
proposals discussed above it will be possible to obtain even stronger constraints
on the non-Newtonian gravity, axionlike particles and on some other related
objects, such as chameleons and symmetrons.

\authorcontributions{Investigation, V.M.M, G.L.K.;  writing, V.M.M, G.L.K.}

\funding{V.M.M.~was partially funded by the Russian Foundation for Basic Research grant
No. 19-02-00453 A.}

\acknowledgments{The authors were partially supported by the Peter
the Great Saint Petersburg Polytechnic University in the framework
of the Program "5--100--2020".
V.M.M.\ was partially supported by the Russian Government Program of Competitive
Growth of Kazan Federal University.}

\conflictsofinterest{The authors declare no conflict of interest.}

\reftitle{References}


\begin{thebibliography}{99}
\bibitem{1}
Kim, J.E.
Light pseudoscalars, particle physics and cosmology.
{\em Phys. Rep.} {\bf 1987}, {\em 150}, 1--177.
\bibitem{2}
Dimopoulos, S.; Giudice, G.F.
Macroscopic forces from supersymmetry.
{\it Phys. Lett. B} {\bf 1996}, {\it 379}, 105--114.
\bibitem{3}
Fujii, Y.
The theoretical background of the fifth force.
{\it Int. J. Mod. Phys. A} {\bf 1991}, {\it 6}, 3505--3557.
\bibitem{4}
Fischbach, E.; Talmadge, C.L.
{\em The Search for Non-Newtonian Gravity}; Springer-Verlag: New
York, USA, 1999.
\bibitem{5}
Weinberg, S.
A New Light Boson?
{\it Phys. Rev. Lett.} {\bf 1978}, {\it 40}, 223--226.
\bibitem{6}
Wilczek, F.
Problem of Strong $P$ and $T$ Invariance in the Presence of Instantons.
{\it Phys. Rev. Lett.} {\bf 1978}, {\it 40}, 279--283.
\bibitem{7}
Kawasaki, M.; Nakayama, K.
Axions: Theory and Cosmological Role.
{\it Annu. Rev. Nucl. Part. Sci.} {\bf 2013}, {\it 63}, 69--95.
\bibitem{8}
Kuzmin, V.A.; Tkachev, I.I.; Shaposhnikov, M.E.
Restrictions imposed on light scalar particles by measurements
of van der Waals forces.
{\it Pis'ma v Zh.\ Eksp.\ Teor.\ Fiz.} {\bf 1982}, {\it 36}, 49--52
[{\it JETP Lett.} {\bf 1982}, {\it 36}, 59--62].
\bibitem{9}
Mostepanenko, V.M.; Sokolov, I.Yu.
The Casimir effect leads to new restrictions on long-range
force constants.
{\it Phys. Lett. A} {\bf 1987}, {\it 125}, 405--408.
\bibitem{10}
Antoniadis, I.; Arkani-Hamed, N.; Dimopoulos, S.; Dvali, G.
New dimensions at a millimeter to a fermi and superstrings at a TeV.
{\em Phys. Lett. B} {\bf 1998}, {\em 436}, 257--263.
\bibitem{11}
Arkani-Hamed, N.; Dimopoulos, S.; Dvali, G.
Phenomenology, astrophysics, and cosmology of theories with
millimeter dimensions and TeV scale quantum gravity.
{\it Phys. Rev. D} {\bf 1999}, {\it 59}, 086004.
\bibitem{12}
Floratos, E.G.; Leontaris, G.K.
Low scale unification, Newton's law and extra dimensions.
{\it Phys. Lett. B} {\bf 1999}, {\it 465}, 95--100.
\bibitem{13}
Kehagias, A.; Sfetsos, K.
Deviations from $1/r^2$ Newton law due to extra dimensions.
{\it Phys. Lett. B} {\bf 2000}, {\it 472}, 39--44.
\bibitem{14}
Ferrer, F.; Nowakowski, M.
Higg- and Goldstone-boson-mediated long range forces.
{\em Phys. Rev. D} {\bf 1999}, {\em 59}, 075009.
\bibitem{15}
Adelberger, E.G.; Fischbach, E.; Krause, D.E.; Newman, R.D.
Constraining the couplings of massive pseudoscalars using gravity
and optical experiments.
{\em Phys. Rev. D} {\bf 2003}, {\em 68}, 062002.
\bibitem{16}
Aldaihan, S.; Krause, D.E.; Long, J.C.; Snow, W.M.
Calculations of the dominant long-range, spin-independent contributions
to the interaction energy between two nonrelativistic Dirac fermions
from double-boson exchange of spin-0 and spin-1 bosons with
spin-dependent couplings.
{\it Phys. Rev. D} {\bf 2017}, {\it 95}, 096005.
\bibitem{17}
Bordag, M.; Klimchitskaya, G.L.; Mohideen, U.; Mostepanenko, V.M.
{\em Advances in the Casimir Effect}; Oxford University
Press: Oxford, UK, 2015.
\bibitem{18}
Smullin, S.J.; Geraci, A.A.; Weld, D.M.; Chiaverini, J.; Holmes, S.;
Kapitulnik, A.
Constraints on Yukawa-type deviations from Newtonian gravity at 20
microns.
{\em Phys. Rev. D} {\bf 2005}, {\em 72}, 122001.
\bibitem{19}
Kapner, D.J.; Cook, T.S.; Adelberger, E.G.; Gundlach, J.H.; Heckel, B.R.;
Hoyle, C.D.; Swanson, H.E.
Tests of the Gravitational Inverse-Square Law below the Dark-Energy
Length Scale.
{\em Phys. Rev. Lett.} {\bf 2007}, {\em 98}, 021101.
\bibitem{20}
Hoskins, J.K.; Newman, R.D.; Spero, R.; Schultz, J.
Experimental tests of the gravitational inverse-square law for mass
separations from 2 to 105 cm.
{\it Phys. Rev. D} {\bf 1985}, {\it 32}, 3084--3095.
\bibitem{21}
Smith, G.L.; Hoyle, C.D.; Gundlach, J.H.; Adelberger, E.G.;
Heckel, B.R.; Swanson, H.E.
Short-range tests of the equivalence principle.
{\it Phys. Rev. D} {\bf 2000}, {\it 61}, 022001.
\bibitem{22}
Schlamminger, S.; Choi, K.-J.; Wagner, T.A.; Gundlach, J.H.;
Adelberger, E.G.
Test of the equivalence principle using a rotating torsion balance.
{\it Phys. Rev. Lett.} {\bf 2008}, {\it 100}, 041101.
\bibitem{23}
Nesvizhevsky, V.V.; Pignol, G.; Protasov, K.V.
Neutron scattering and extra short range interactions.
{\it Phys. Rev. D} {\bf 2008}, {\it 77}, 034020.
\bibitem{24}
Kamiya, Y.; Itagami, K.; Tani, M.; Kim, G.N.; Komamiya, S.
Constraints on New Gravitylike Forces in the Nanometer Range.
{\it Phys. Rev. Lett.} {\bf 2015}, {\it 114}, 161101.
\bibitem{25}
Haddock, C.C.; Oi, N.; Hirota, K.; Ino, T.; Kitaguchi, M.; Matsumoto, S.;
Mishima, K.; Shima, T.; Shimizu, H.M.; Snow, W.M.; Yoshioka, T.
Search for deviations from the inverse square law of gravity at nm range
using a pulsed neutron beam.
{\it Phys. Rev. D} {\bf 2018}, {\it 97}, 062002.
\bibitem{26}
Chen, Y.J.; Tham, W.K.; Krause, D.E.; L\'{o}pez, D.; Fischbach, E.; Decca, R.S.
Stronger Limits on Hypothetical Yukawa Interactions in the
30--8000 Nm Range.
{\em Phys. Rev. Lett.} {\bf 2016}, {\em 116}, 221102.
\bibitem{27}
Kim, J.E.
Weak-Interaction Singlet and Strong CP Invariance.
{\em Phys. Rev. Lett.} {\bf 1979}, {\em 43}, 103--107.
\bibitem{28}
Shifman, M.A.; Vainstein, A.I.; Zakharov, V.I.
Can confinement ensure natural CP invariance of strong interactions?
{\it Nucl. Phys. B} {\bf 1980}, {\it 166}, 493--506.
\bibitem{29}
Zhitnitskii, A.P.
On the possible suppression of axion-hadron interactions.
{\it Sov. J. Nucl. Phys.} {\bf 1980}, {\it 31}, 260--263.
\bibitem{30}
Dine, M.; Fischler, F.; Srednicki, M.
A simple solution to the strong CP problem with a harmless axion.
{\it Phys. Lett. B} {\bf 1981}, {\it 104}, 199--202.
\bibitem{31}
Rosenberg, L.J.; van\ Bibber, K.A.
Searches for invisible axions.
{\it Phys. Rep.} {\bf 2000}, {\it 325}, 1--39.
\bibitem{32}
Raffelt, G.G.
Axions -- motivation, limits and searches.
{\it J. Phys. A: Math. Theor.} {\bf 2007}, {\it 40}, 6607--6620.
\bibitem{33}
Ivastorza, I.G.; Redondo, J.
New experimental approaches in the search for axion-like particles.
{\it Progr. Part. Nucl. Phys.} {\bf 2018}, {\it 102}, 89--159.
\bibitem{34}
Raffelt, G.
Limits on a $CP$-violating scalar axion-nucleon interaction.
{\em Phys. Rev. D} {\bf 2012}, {\em 86}, 015001.
\bibitem{35}
Vasilakis, G.; Brown, J.M.; Kornak, T.R.; Romalis, M.V.
Limits on New Long Range Nuclear Spin-Dependent Forces Set
with a K-${}^3$He Comagnetometer.
{\it Phys. Rev. Lett.} {\bf 2009}, {\it 103}, 261801.
\bibitem{36}
Adelberger, E.G.; Heckel, B.R.; Hoedl, S.; Hoyle, C.D.; Kapner, D.J.;
Upadhye, A.
Particle-Physics Implications of a Recent Test of the Gravitational
Inverse-Square Law.
{\em Phys. Rev. Lett.} {\bf 2007}, {\em 98}, 131104.
\bibitem{37}
Klimchitskaya, G.L.; Mostepanenko, V.M.
Improved constraints on the coupling constants of axion-like
particles to nucleons from recent Casimir-less experiment.
{\em Eur. Phys. J. C} {\bf 2015}, {\em 75}, 164.
\bibitem{38}
Ramsey, N.F.
The tensor force between two protons at long range.
{\it Physica A} {\bf 1979}, {\it 96}, 285--289.
\bibitem{39}
Ledbetter, M.P.; Romalis, M.V.; Jackson Kimball, D.F.
Constraints on Short-Range Spin-Dependent Interactions from Scalar
Spin-Spin Coupling in Deuterated Molecular Hydrogen.
{\it Phys. Rev. Lett.} {\bf 2013}, {\it 110}, 040402.
\bibitem{39a}
{  Antoniadis, I.; Baessler, S.; B\"{u}cher, M.; Fedorov, V.\ V.;
Hoedl, S.; Lambrecht, A.; Nesvizhevsky, V.\ V.; Pignol, G.;  Protasov, K.\ V.;
Reynaud, S.; Sobolev, Yu.
Short-range fundamental forces.
{\it Compt. Rend.} {\bf 2011}, {\it 12}, 755--778.}
\bibitem{40}
Kardar, M.; Golestanian, R.
The "friction" of vacuum, and other fluctuation-induced forces.
{\it Rev. Mod. Phys.} {\bf 1999}, {\it 71}, 1233--1245.
\bibitem{41}
Chen, F.; Mohideen, U.; Klimchitskaya, G.L.; Mostepanenko, V.M.
Demonstration of the Lateral Casimir Force.
{\it Phys. Rev. Lett.} {\bf 2002}, {\it 88}, 101801.
\bibitem{42}
Chen, F.; Mohideen, U.; Klimchitskaya, G.L.; Mostepanenko, V.M.
Experimental and theoretical investigation of the lateral
Casimir force between corrugated surfaces.
{\it Phys. Rev. A} {\bf 2002}, {\it 66}, 032113.
\bibitem{43}
Chiu, H.C.; Klimchitskaya, G.L.; Marachevsky, V.N.; Mostepanenko, V.M.;
Mohideen, U.
Demonstration of the asymmetric lateral Casimir force between
corrugated surfaces in the nonadditive regime.
{\em Phys. Rev. B} {\bf 2009}, {\em 80}, 121402(R).
\bibitem{44}
Chiu, H.C.; Klimchitskaya, G.L.; Marachevsky, V.N.; Mostepanenko, V.M.;
Mohideen, U.
Lateral Casimir force between sinusoidally corrugated surfaces:
Asymmetric profiles, deviations from the proximity force approximation, and
comparison with exact theory.
{\em Phys. Rev. B} {\bf 2010}, {\em 81}, 115417.
\bibitem{45}
Bezerra, V.B.; Klimchitskaya, G.L.; Mostepanenko, V.M.; Romero, C.
Advance and prospects in constraining the Yukawa-type corrections to
Newtonian gravity from the Casimir effect.
{\it Phys. Rev. D} {\bf 2010}, {\it 81}, 055003.
\bibitem{46}
Mostepanenko, V.M.; Novello, M.
Constraints on non-Newtonian gravity from the Casimir force
measurements between two crossed cylinders.
{\it Phys. Rev. D} {\bf 2001}, {\it 63}, 115003.
\bibitem{47}
Ederth, T.
Template-stripped gold surfaces with 0.4-nm rms roughness suitable
for force measurements: Application to the Casimir force in the
20--100-nm range.
{\it Phys. Rev. A} {\bf 2000}, {\it 62}, 062104.
\bibitem{48}
Klimchitskaya, G.L.; Mohideen, U.; Mostepanenko, V.M.
The Casimir force between real materials: Experiment and theory.
{\em Rev. Mod. Phys.} {\bf 2009}, {\em 81}, 1827--1885.
\bibitem{49}
Banishev, A.A.; Wagner, J.; Emig, T.; Zandi, R.; Mohideen, U.
Demonstration of Angle-Dependent Casimir Force between Corrugations,
{\it Phys. Rev. Lett.} {\bf 2013}, {\it 110}, 250403.
\bibitem{50}
Banishev, A.A.; Wagner, J.; Emig, T.; Zandi, R.; Mohideen, U.
Experimental and theoretical investigation of the angular dependence of
the Casimir force between sinusoidally corrugated surfaces.
{\it Phys. Rev. B} {\bf 2014}, {\it 89}, 235436.
\bibitem{51}
Fosco, C.D.; Lombardo, F.C.; Mazzitelli, F.D.
Proximity force approximation for the Casimir energy as a derivative expansion.
{\it Phys. Rev. D} {\bf 2011}, {\it 84}, 105031.
\bibitem{52}
Bimonte, G.; Emig, T.; Kardar, M.
Material dependence of Casimir forces: Gradient expansion beyond proximity.
{\it Appl. Phys. Lett.} {\bf 2012}, {\it 100}, 074110.
\bibitem{53}
Klimchitskaya, G.L.; Mohideen, U.; Mostepanenko, V.M.
Constraints on corrections to Newtonian gravity from two recent
measurements of the Casimir interaction between metallic surfaces.
{\it Phys. Rev. D} {\bf 2013}, {\it 87}, 125031.
\bibitem{54}
Mostepanenko, V.M.
Progress in constraining axion and non-Newtonian gravity from the
Casimir effect.
{\it Int J. Mod. Phys. A} {\bf 2016}, {\it 31}, 1641020.
\bibitem{55}
Decca, R.S.; L\'{o}pez, D.; Fischbach, E.; Klimchitskaya, G.L.; Krause, D.E.;
Mostepanenko, V.M.
Tests of new physics from precise measurements of the Casimir
pressure between two gold-coated plates.
{\em Phys. Rev. D} {\bf 2007}, {\em 75}, 077101.
\bibitem{56}
Decca, R.S.; L\'{o}pez, D.; Fischbach, E.; Klimchitskaya, G.L.; Krause, D.E.;
Mostepanenko, V.M.
Novel constraints on light elementary particles and
extra-dimensional physics from the Casimir effect.
{\em Eur. Phys. J. C} {\bf 2007}, {\em 51}, 963--975.
\bibitem{57}
Bimonte, G.
Going beyond PFA: A precise formula for the sphere-plate Casimir force.
{\it Europhys. Lett.} {\bf 2017}, {\em 118}, 20002.
\bibitem{58}
Hartmann, M.; Ingold, G.-L.; Maia Neto, P.A.
Plasma versus Drude Modeling of the Casimir Force: Beyond the Proximity
Force Approximation.
{\em Phys. Rev. Lett.} {\bf 2017}, {\em 119}, 043901.
\bibitem{59}
Mostepanenko, V.M.; Bezerra, V.B.; Decca, R.S.; Geyer, B.; Fischbach, E.;
Klimchitskaya, G.L.; Krause, D.E.; L\'{o}pez, D.; Romero, C.
Present status of controversies regarding the thermal Casimir force.
{\it J. Phys.: Math. Gen.} {\bf 2006}, {\em 39}, 6589--6600.
\bibitem{60}
Bimonte, G.; L\'{o}pez, D.; Decca, R.S.
Isoelectronic determination of the thermal Casimir force.
{\em Phys. Rev. B} {\bf 2016}, {\em 93}, 184434.
\bibitem{60a}
{  Mostepanenko, V.M.; Bezerra, V.B.; Klimchitskaya, G.L.; Romero, C.
New constraints on the Yukawa-type interactions from the Casimir effect.
{\it Int. J. Mod. Phys. A} {\bf 2012}, {\it 27}, 1260015.}
\bibitem{61}
Decca, R.S.; L\'{o}pez, D.; Chan, H.B.; Fischbach, E.; Krause, D.E.; Jamell, C.R.
Constraining New Forces in the Casimir Regime Using the
Isoelectronic Technique.
{\em Phys. Rev. Lett.} {\bf 2005}, {\em 94}, 240401.
\bibitem{62}
Wang, J.; Guan, S.; Chen, K.; Wu, W.; Tian, Z.; Luo, P.; Jin, A.; Yang, S.;
Shao, C.; Luo, J.
Test of non-Newtonian gravitational forces at micrometer range with
two-dimensional force mapping.
{\em Phys. Rev. D} {\bf 2016}, {\em 94}, 122005.
\bibitem{63}
Masuda, M.; Sasaki, M.
Limits on Nonstandard Forces in the Submicrometer Range.
{\em Phys. Rev. Lett.} {\bf 2009}, {\em 102}, 171101.
\bibitem{64}
Bohr, A.; Mottelson, B.R.
{\it Nuclear Structure, Vol.~1}; Benjamin: New York, USA, 1969.
\bibitem{65}
Bezerra, V.B.; Klimchitskaya, G.L.; Mostepanenko, V.M.; Romero, C.
Constraining axion coupling constants from measuring the Casimir
interaction between polarized test bodies.
{\em Phys. Rev. D} {\bf 2016}, {\em 94}, 035011.
\bibitem{66}
Drell, S.D.; Huang, K.
Many-Body Forces and Nuclear Saturation.
{\it Phys. Rev.} {\bf 1953}, {\em 91}, 1527--1542.
\bibitem{67}
Bezerra, V.B.; Klimchitskaya, G.L.; Mostepanenko, V.M.; Romero, C.
Constraining axion-nucleon coupling constants from measurements of
effective Casimir pressure by means of micromachined oscillator.
{\em Eur. Phys. J. C} {\bf 2014}, {\em 74}, 2859.
\bibitem{68}
Bezerra, V.B.; Klimchitskaya, G.L.; Mostepanenko, V.M.; Romero, C.
Constraints on axion-nucleon coupling constants from measuring the
Casimir force between corrugated surfaces.
{\em Phys. Rev. D} {\bf 2014}, {\em 90}, 055013.
\bibitem{69}
Klimchitskaya, G.L.; Mostepanenko, V.M.
Constraints on axionlike particles and non-Newtonian gravity from
measuring the difference of Casimir forces.
{\em Phys. Rev. D} {\bf 2017}, {\em 95}, 123013.
\bibitem{70}
Bezerra, V.B.; Klimchitskaya, G.L.; Mostepanenko, V.M.; Romero, C.
Stronger constraints on an axion from measuring the Casimir
interaction by means of a dynamic atomic force microscope.
{\em Phys. Rev. D} {\bf 2014}, {\em 89}, 075002.
\bibitem{71}
Chang, C.C.; Banishev, A.A.; {Castillo-Garza}, R.; Klimchitskaya, G.L.;
Mostepanenko, V.M.; Mohideen, U.
Gradient of the Casimir force between Au surfaces of a sphere and a
plate measured using an atomic force microscope in a frequency-shift
technique.
{\em Phys. Rev. B} {\bf 2012}, {\em 85}, 165443.
\bibitem{72}
Bezerra, V.B.; Klimchitskaya, G.L.; Mostepanenko, V.M.; Romero, C.
Constraints on the parameters of an axion from measurements of the
thermal Casimir-Polder force.
{\em Phys. Rev. D} {\bf 2014}, {\em 89}, 035010.
\bibitem{73}
Obrecht, J.M.; Wild, R.J.; Antezza, M.; Pitaevskii, L.P.; Stringari, S.;
Cornell, E.A.
Measurement of the Temperature Dependence of the Casimir-Polder Force.
{\em Phys. Rev. Lett.} {\bf 2007}, {\em 98}, 063201.
\bibitem{74}
Klimchitskaya, G.L.; Kuusk, P.; Mostepanenko, V.M.
Constraints on non-Newtonian gravity and axionlike particles from
measuring the Casimir force in nanometer separation range.
{\em Phys. Rev. D} {\bf 2020}, {\em 101}, 056013.
\bibitem{75}
Sedighi, M.; Svetovoy, V.B.; Palasantzas, G.
Casimir force measurements from silicon carbide surfaces.
{\it Phys. Rev. B} {\bf 2016}, {\it 93}, 085434.
\bibitem{76}
Sedighi, M.; Svetovoy, V.B.; Broer, W.H.; Palasantzas, G.
Casimir forces from conductive silicon carbide surfaces.
{\it Phys. Rev. B} {\bf 2014}, {\it 89}, 195440.
\bibitem{77}
Svetovoy, V.B.; van\ Zwol, P.J.; Palasantzas, G.;
De\ Hosson, Th.M.
Optical properties of gold films and the Casimir force.
{\it Phys. Rev. B} {\bf 2008}, {\it 77}, 035439.
\bibitem{78}
Tan, W.-H.; Du, A.-B.; Dong, W.-C.; Yang, S.-Q.; Shao, C.-G.;
Guan, S.-G.; Wang, Q.-L.; Zhan, B.-F.; Luo, P.-S.; Tu, L.-C.; Luo, J.
Improvement for Testing the Gravitational Inverse-Square Law at the
Submillimeter Range.
{\it Phys. Rev. Lett.} {\bf 2020}, {\it 124}, 051301.
\bibitem{79}
Klimchitskaya, G.L.
Recent breakthrough and outlook in constraining the non-Newtonian
gravity and axion-like particles from Casimir physics.
{\em Eur. Phys. J. C} {\bf 2017}, {\em 77}, 315.
\bibitem{80}
Almasi, A.; Brax, P.; Iannuzzi, D.; Sedmik, R.I.P.
Force sensor for chameleon and Casimir force experiments with
parallel-plate configuration.
{\em Phys. Rev. D} {\bf 2015}, {\em 91}, 102002.
\bibitem{81}
Sedmik, R.; Brax, P.
Status Report and first Light from Cannex: Casimir Force
Measurements between flat parallel Plates.
{\em J. Phys.: Conf. Ser.} {\bf 2018}, {\em 1138}, 012014.
\bibitem{82}
Klimchitskaya, G.L.; Mostepanenko, V.M.; Sedmik, R.I.P.;
Abele, H.
Prospects for Searching Thermal Effects, Non-Newtonian Gravity and
Axion-Like Particles: CANNEX Test of the Quantum Vacuum.
{\it Symmetry} {\bf 2019}, {\it 11}, 407.
\bibitem{83}
Klimchitskaya, G.L.; Mostepanenko, V.M.; Sedmik, R.I.P.
Casimir pressure between metallic plates out of thermal equilibrium:
Proposed test for the relaxation properties of free electrons.
{\it Phys. Rev. A} {\bf 2019}, {\it 100}, 022511.
\bibitem{84}
Bennett, R.; O'Dell, D.H.J.
Revealing short-range non-Newtonian gravity through Casimir-Polder
shielding.
{\it New J. Phys.} {\bf 2019}, {\it 21}, 033032.
\bibitem{85}
Borkowski, M.; Buchachenko, A.A.; Ciury{\l}o, R.; Julienne, P.S.;
Yamada, H; Kikuchi, Y.; Takasu, Y.; Takahashi, Y.
Weakly bound molecules as sensors of new gravitylike forces.
{\it Sci. Rep.} {\bf 2019}, {\it 9}, 14807.
\bibitem{86}
Safronova, M.S.; Budker, D.; DeMille, D.; Jackson Kimball, D.F.;
Derevianko, A.; Clark, C.W.
Search for new physics with atoms and molecules.
{\it Rev. Mod. Phys.} {\bf 2018}, {\it 90}, 025008.
\bibitem{87}
Liu, J.; Zhu, K.-D.
Detecting large extra dimensions with optomechanical levitated sensors.
{\it Eur. Phys. J. C} {\bf 2019}, {\it 79}, 18.
\bibitem{88}
Sedmik, R.I.P.
Casimir and non-Newtonian force experiment (CANNEX): Review,
status, and outlook.
{\it Int. J. Mod. Phys. A} {\bf 2020}, {\it 35}, 2040008.
\bibitem{89}
Elder, B.; Vardanyan, V.; Arkami, Y.; Brax, P.; Davis, A.-C.;
Decca, R.S.
Classical symmetron force in Casimir experiments.
{\it Phys. Rev. D} {\bf 2020}, {\it 101}, 064065.
\end{thebibliography}
\end{document}